\documentclass[twocolumn,nopacs,prb,amsfonts,amsmath,amssymb,floatfix]{revtex4} 
\usepackage{color}
\usepackage{hhline}	
\usepackage{mathrsfs}
\usepackage{graphicx}
\usepackage{dcolumn}
\usepackage{bm}
\usepackage{multirow}
\usepackage{booktabs}
\usepackage{afterpage}
\usepackage{amsmath}
\usepackage{braket}

\begin{document}

\title{Prediction of the High Thermoelectric Performance of Pnictogen Dichalcogenide Layered Compounds with Quasi-One-Dimensional Gapped Dirac-like Band Dispersion}

\author{Masayuki Ochi}
\author{Hidetomo Usui}
\author{Kazuhiko Kuroki}
\affiliation{Department of Physics, Osaka University, Machikaneyama-cho, Toyonaka, Osaka 560-0043, Japan}

\date{\today}
\begin{abstract}
Thermoelectric power generation has been recognized as one of the most important technologies, and high-performance thermoelectric materials have long been pursued. However, because of the large number of candidate materials, this quest is extremely challenging, and it has become clear that a firm theoretical concept from the viewpoint of band-structure engineering is needed.
In this study, we theoretically demonstrate that pnictogen-dichalcogenide layered compounds, which originally attracted attention as a family of superconductors and have recently been investigated as thermoelectric materials, can exhibit very high thermoelectric performance with elemental substitution.
In particular, we clarify a promising guiding principle for materials design and find that LaOAsSe$_2$, a material that has yet to be synthesized, has a powerfactor that is six times as large as that of the known compound LaOBiS$_2$ and can exhibit a very large $ZT$ under some plausible assumptions.
This large enhancement of the thermoelectric performance originates from the quasi-one-dimensional gapped Dirac-like band dispersion, which is realized by the square-lattice network.
Our study offers one ideal limit of the band structure for thermoelectric materials.
Because our target materials have high controllability of constituent elements and feasibility of carrier doping, experimental studies along this line are strongly awaited.
\end{abstract}

\maketitle

\section{Introduction}

Exploring high-performance thermoelectric materials is of crucial importance for the efficient use of renewable heat energy.
The efficiency of thermoelectric conversion is evaluated by the dimensionless figure of merit, $ZT=\sigma S^2 T \kappa^{-1}$, where $\sigma$, $S$, $T$, and $\kappa$ are the electrical conductivity, Seebeck coefficient, temperature, and thermal conductivity, respectively.
A considerable number of studies have been conducted to increase the $ZT$ value, e.g., by nanostructuring~\cite{nano1,nano2,nano3} and by the search for appropriate materials.
Although the great importance of this quest has been widely recognized, it is extremely challenging to find favorable materials among the large number of candidates.
For this purpose, a firm theoretical concept based on band-structure engineering is needed.
It is also important for candidate materials to have large degrees of freedom, such as for elemental substitution and carrier doping, because of the high sensitivity of the thermoelectric performance to the band structure in a narrow energy window around the chemical potential, which often requires fine-tuning of the band structure and carrier concentration to maximize the performance.

One promising strategy is to reduce the dimensionality of the electronic structure because a low-dimensional electronic structure can simultaneously host a large density of states (DOS) and high group velocity along specific directions, both of which are advantageous for high-thermoelectric performance~\cite{Hicksone,Hicksone2,usuione}.
These studies noted that the thermal conductivity is also suppressed for low-dimensional materials.
Indeed, some high-performance thermoelectric materials, such as Bi$_2$Te$_3$ alloys~\cite{nano3,Bi2Te3_rev}, BiCuSeO~\cite{BiCuSeO}, and Na$_x$CoO$_2$~\cite{NaxCOO2}, have layered structures.
Recent findings of high-thermoelectric performance in (quasi-)one-dimensional materials such as silicon nanowires~\cite{nano1}, carbon nanotubes~\cite{CNT1,CNT2}, In$_4$Se$_{3-\delta}$~\cite{In4Se3}, SbCrSe$_3$~\cite{SbCrSe3}, and Ta$_4$SiTe$_4$~\cite{Ta4SiTe4}, are also encouraging.
Along this line, one promising candidate is pnictogen-dichalcogenide layered compounds~\cite{BiS2,BiS2_2,BiS2review,BiS2review2,BiS2review3}, which originally attracted attention as a family of layered superconductors and have recently been observed to exhibit favorable thermoelectric performance~\cite{BiS2_review_thermo}.
Although many studies have focused on their superconducting properties~\cite{BiS2review,BiS2review2,BiS2review3,BiS2_corentin,BiS2_suzuki}, there have thus far been few experimental~\cite{LaOFBiS2_thermo,LaOFBiS2_Hall,LaOBiSSe_1,LaOBiSSe_HP,LaOBiSSe_2,LaOBiSSe_3,EuFBiS2,LaPbBiS3O} or theoretical~\cite{BiS2_thermo_theory, BiS2_thermo_theory2} investigations on their thermoelectric properties.
Nevertheless, a relatively high $ZT$ value $\sim$ 0.36 with a low thermal conductivity $\kappa \sim 1$ W m$^{-1}$ K$^{-1}$ has already been reported for LaOBiSSe at $\sim$ 650 K~\cite{LaOBiSSe_HP}.
Although this $ZT$ value is not very high compared with those of materials currently used in industry, the rich variety of constituent elements of pnictogen-dichalcogenide layered compounds~\cite{BiS2review,BiS2review2,BiS2review3}, a large part of which remains unexplored, offers a promising opportunity for improving their thermoelectric performance.
Moreover, some theoretical studies have shown that a quasi-one-dimensional band structure is realized in these compounds~\cite{Usui}, which can be a promising platform for high-thermoelectric performance.

In this study, we uncovered a promising guiding principle for improving the thermoelectric performance of pnictogen-dichalcogenide layered compounds by careful theoretical analysis.
In particular, we discovered that a material that has yet to be synthesized, LaOAsSe$_2$, can exhibit distinguished thermoelectric performance.
The square lattice therein offers an ideal arena for the quasi-one-dimensional gapped Dirac-like band dispersion and therefore for high-thermoelectric performance.
This study offers not only an important clue for optimizing the thermoelectric performance of pnictogen-dichalcogenide layered compounds but also a valuable example of {\it materials design} to improve their functionality, which has received increased attention today as a result of the rapid development in materials informatics.

This paper is organized as follows. Section~\ref{sec:cal} presents a detailed procedure of our calculation and a brief description of the Boltzmann transport theory used in our study.
We present the basic thermoelectric properties of a prototypical pnictogen-dichalcogenide compound, LaOBiS$_2$, in Sec.~\ref{sec:lao}.
Our theoretical analysis of several factors that dominate the thermoelectric performance of Bi-chalcogenide layered compounds is presented in Sec.~\ref{sec:key}.
Based on the guiding principle for improving their performance, which is observed in Sec.~\ref{sec:key}, we compare the thermoelectric performance of some compositions in Sec.~\ref{sec:material}.
Additional discussion on the ideal condition for the band structure we found is presented in Sec.~\ref{sec:disc}. Section~\ref{sec:sum} is devoted to the conclusion of this study.

\section{Methods of calculation\label{sec:cal}}

For first-principles band structure calculation, we used the modified Becke--Johnson (mBJ) potential proposed by Tran and Blaha~\cite{mBJ1,mBJ2} and the full-potential linearized augmented plane-wave method, as implemented in the \textsc{wien2k} code~\cite{wien2k}.
The experimental crystal structure of LaOBiS$_2$ was taken from Ref.~[\onlinecite{TetraStruct}] ($a=4.05$ \AA, $c=13.74$ \AA).
Although slight symmetry reduction was experimentally observed~\cite{SymLow}, we assumed the tetragonal crystal structure (space group: $P$4/$nmm$) for simplicity. This treatment suffices for our aim of obtaining a theoretical guiding principle for materials design of pnictogen-dichalcogenide layered compounds, many of which have the tetragonal symmetry. 
Short discussions regarding the symmetry reduction of the crystal structure are presented in the APPENDIX.
We also took the experimental crystal structures of NdOBiS$_2$ ($a=4.00$ \AA, $c=13.46$ \AA) and NdOSbS$_2$ ($a=3.98$ \AA, $c=13.80$ \AA) from Refs.~\cite{NdOBiS2_strct} and \cite{NdOSbS2_strct}, respectively~\cite{note_NdOBiS2_strct}, both of which belong to the space group $P$4/$nmm$.
To represent the strongly localized 4$f$ orbitals of Nd atoms, we employed the open-core treatment where the 4$f^3$ states are included into the core states, i.e., not explicitly treated as the valence states.
The $RK_{\rm max}$ parameter was set to 8.00.
Because LaOAsSe$_2$ has yet to be synthesized to our knowledge, we determined its crystal structure through structural optimization using the projector augmented wave method~\cite{paw} and the Perdew--Burke--Ernzerhof parameterization of the generalized gradient approximation (PBE-GGA)~\cite{PBE} as implemented in the \textsc{vasp} code~\cite{vasp1,vasp2,vasp3,vasp4}. 
For the structural optimization, we employed a plane-wave cutoff energy of 600 eV, a $24\times 24\times 6$ $\bm{k}$-mesh, and the convergence criterion for the Hellmann--Feynman force on each atom of 0.01 eV/\AA\ without the inclusion of the spin--orbit coupling (SOC). We also assumed the space group $P$4/$nmm$ there.
The obtained lattice parameters were $a=3.99$ \AA\ and $c=14.33$ \AA.

After the first-principles band structure calculation, we extracted the Wannier functions from the calculated band structures using the \textsc{Wien2Wannier} and \textsc{Wannier90} codes~\cite{Wannier1,Wannier2,Wannier90,Wien2Wannier}.
In this study, we took the $p_{x,y,z}$ orbitals of all the Bi (Sb, As), S (Se), and O atoms as the Wannier basis set.
Here, we considered both the valence and conduction band dispersions by accounting for these atomic orbitals while we concentrated on the electron-carrier doping, which is feasible in experiments. This is because the bipolar conduction can have a sizable contribution to the transport properties at high temperatures for small-band-gap materials, a category to which our target materials can belong, as we shall see later in this paper.
We did not perform the maximal localization procedure for the Wannier functions to prevent orbital mixing among the different spin components and to allow for a more intuitive understanding of the hopping parameters.
We used an $8\times 8\times 8$ $\bm{k}$-mesh for constructing the Wannier functions.
Then, we constructed the tight-binding model with the obtained hopping parameters among the Wannier functions and analyzed the transport properties using this model.
For this purpose, we employed Boltzmann transport theory~\cite{Boltz}, where the transport coefficients ${\bf K}_{\nu}$ are represented as follows:
\begin{align}
{\bf K}_\nu= \tau\sum_{n,{\bm{k}}} \bm{v}_{n,\bm{k}}\otimes\bm{v}_{n,\bm{k}}\left[-\frac{\partial f_0}{\partial \epsilon_{n,\bm{k}}}\right](\epsilon_{n,\bm{k}}-\mu(T))^\nu ,\label{eq:transp}
\end{align}
by using the Fermi--Dirac distribution function $f_0$, chemical potential $\mu(T)$, energy $\epsilon_{n,\bm{k}}$ and group velocity $\bm{v}_{n,\bm{k}}$ of the one-electron orbital on the $n$-th band at some $\bm{k}$-point and the relaxation time $\tau$, which was assumed to be constant in this study.
Here, $\mu(T)$ was determined to provide a given carrier density against the temperature change for calculations with the fixed carrier density.
The electrical conductivity ${\boldsymbol \sigma}$, Seebeck coefficient ${\bf S}$, and electrical thermal conductivity ${\boldsymbol \kappa}_{\rm el}$ are expressed as follows:
\begin{align}
{\boldsymbol \sigma}=e^2{\bf K}_0,\ \ \  {\bf S}=-\frac{1}{eT}{\bf K}_0^{-1}{\bf K}_1,\\ 
{\boldsymbol \kappa}_{\rm el}=\frac{1}{T}\left[{\bf K}_2-{\bf K}_1{\bf K}_0^{-1}{\bf K}_1 \right] ,
\end{align}
where $e$ ($>0$) is the elementary charge. The powerfactor PF is defined as PF $=\sigma S^2$ for the diagonal components of these tensors.
For the transport calculations, we used our own code.
Note that the tetragonal crystal structure, on which we focused here, forces the off-diagonal components of the transport coefficients to be zero.
We assumed the thermal conductivity can be represented as the sum of the electrical thermal conductivity ${\boldsymbol \kappa}_{\rm el}$ and the lattice electrical thermal conductivity ${\boldsymbol \kappa}_{\rm lat}$, namely, ${\boldsymbol \kappa} = {\boldsymbol \kappa}_{\rm el}+{\boldsymbol \kappa}_{\rm lat}$.
To simulate the carrier doping, we adopted the rigid band approximation, as it has been validated by theoretical observation that the electron-carrier doping does not have a large effect on the band structure of LaOBiS$_2$~\cite{rigid_band}.
In this study, we investigated the transport properties at $T=300$ K unless noted otherwise.
To achieve sufficient convergence, we employed a $240\times 240\times 60$ $\bm{k}$-mesh for calculating the transport coefficients, whereas a $480\times 480\times 60$ $\bm{k}$-mesh was used only for evaluating the $ZT$ values in Sec.~\ref{sec:material}, where a lower-temperature region was also investigated.
We note that the phonon drag effect, which is beyond the scope of our present study, can affect the thermoelectric properties in the low-temperature regime, whereas it is expected to be suppressed in the high-temperature regime where the $ZT$ value increases.
All the results, except those in Figures 1(a)--(b), 2(c), and the black dotted lines in Figures 1(c), 2(a), and 5(k), were calculated using the tight-binding model, and therefore, the $4f$ bands do not appear there.

\section{Results and Discussion\label{sec:res}}

\subsection{Thermoelectric properties of LaOBiS$_2$\label{sec:lao}}

\begin{figure}
\begin{center}
\includegraphics[width=8.5 cm]{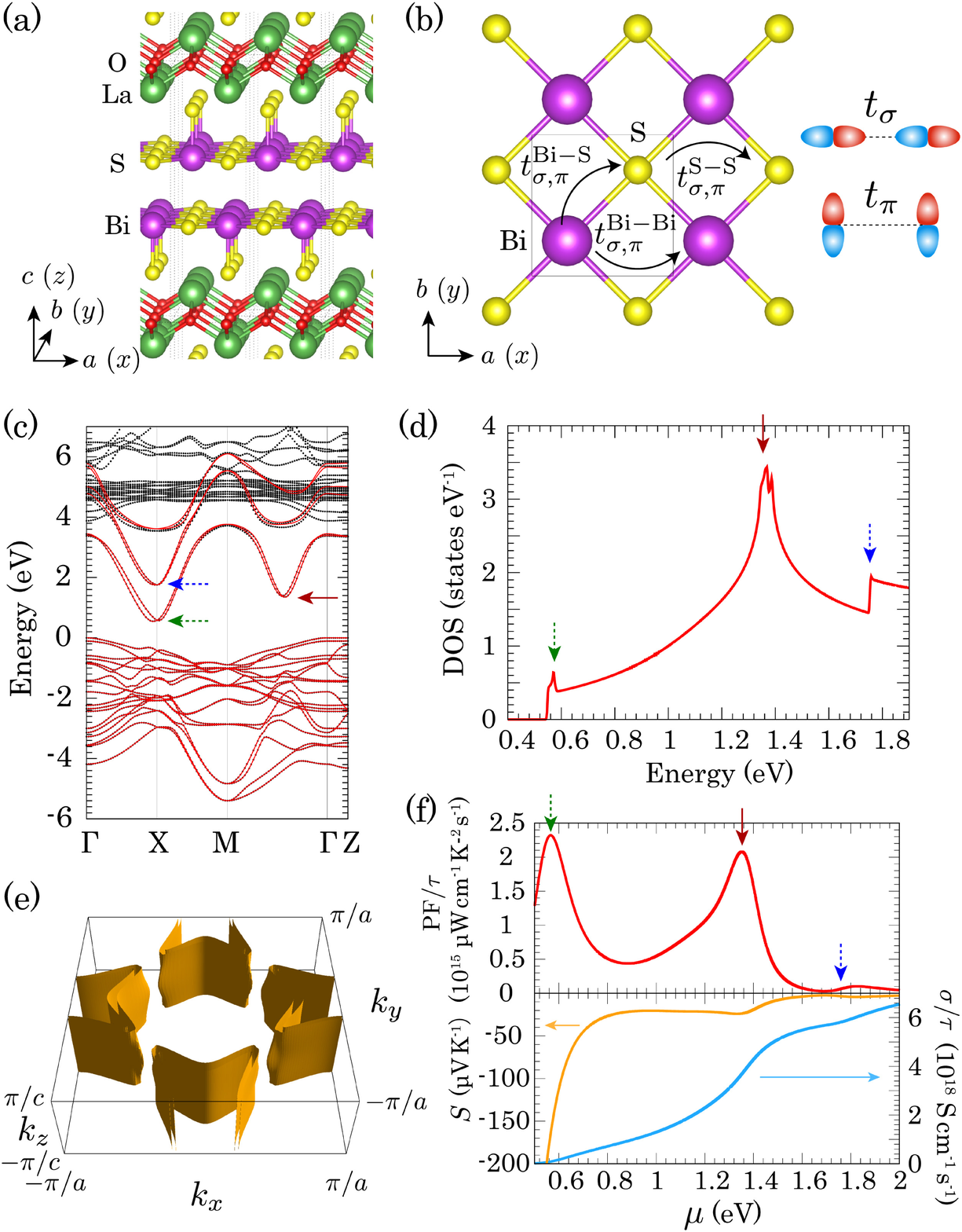}
\caption{(a) Crystal structure of tetragonal LaOBiS$_2$ and (b) the square lattice therein with the definition of some hopping amplitudes.
The crystal structures were depicted using \textsc{VESTA} software~\cite{VESTA}.
(c) Electronic band structures obtained using first-principles calculations (black dotted lines) and model calculations (red solid lines).
The energy of the valence band maximum was set to zero.
(d) DOS, (e) Fermi surface at $\mu=$1.34 eV, which is close to the vHs, and (f) transport properties calculated using the tight-binding model.
The three arrows in the upper half of (f) correspond to those shown in (c) and (d).
All the calculations in this figure include the SOC.}
 \label{fig:1}
 \end{center}
\end{figure}

Figure~\ref{fig:1} presents the crystal structure, basic electronic structure, and transport properties of LaOBiS$_2$, which we adopted here as a typical pnictogen-dichalcogenide layered compound. First, we briefly review the basic electronic structure of LaOBiS$_2$ but refer the readers to several review articles~\cite{BiS2review,BiS2review2,BiS2review3} for more details.
As observed in Figure~\ref{fig:1}(a)--(b), the square lattice consisting of Bi and S atoms and the blocking layers consisting of La and O atoms are alternatively stacked. 
In Figure~\ref{fig:1}(c), the band structure calculated using the tight-binding model with the spin--orbit coupling is marked with red solid lines together with the first-principles marked with black dotted lines. It is apparent that our tight-binding model well reproduces the first-principles band structure.
The band gap calculated using the mBJ potential, 0.6 eV, is fairly consistent with the experimental ones, 0.7~\cite{LaOFBiS2_gap2}, 0.8~\cite{LaOFBiS2_gap1}, and 1.0~\cite{LaOBiS2_gap} with different rates of F substitution and different measurement methods, whereas the popular PBE-GGA calculation provides a greatly underestimated value of 0.2 eV~\cite{note:gap}.

Figure~\ref{fig:1}(d) shows the DOS for the conduction bands, which consist of Bi and S states.
Here, we focus on the conduction band dispersion because the electron carriers are doped into pnictogen-dichalcogenide layered compounds, e.g., by (partial) substitution of F atoms for O atoms in experiments.
Three characteristic DOS peaks are indicated by arrows in this figure. The two arrows with dotted lines at the energy levels of approximately 0.6 and 1.8 eV indicate the two band edges around the X point, and the most prominent DOS peak at the energy level of approximately 1.4 eV corresponds to the van Hove singularity (vHs).
The Fermi surface at the energy level near the vHs is shown in Figure~\ref{fig:1}(e).
As observed in this figure, the conduction band dispersion is almost flat along the $z$-direction because of the existence of the blocking layers.
This feature can also be verified through the Fermi surfaces at different energy levels that were shown in previous studies, e.g., Ref.~\onlinecite{FS_BiS2}.
We note that the $\bm{k}$-points corresponding to the vHs are positioned roughly around $\bm{k}=(\pm \pi/2, \pm \pi/2)$.
It is also characteristic that the band splitting induced by the inter-BiS$_2$-layer coupling is small~\cite{Usui,BiS2_HP,sym_akashi}, which enhances the thermoelectric performance of these materials by increasing the DOS through the weak degeneracy for two BiS$_2$ layers.

Next, we move on to the thermoelectric properties.
The calculated PF$/ \tau$, $S$, and $\sigma$/$\tau$ are presented in Figure~\ref{fig:1}(f) as a function of the chemical potential $\mu$. We only show the in-plane diagonal components of the transport quantities, e.g., $S=S_{xx}=S_{yy}$, throughout this paper because the off-diagonal elements vanish under the tetragonal symmetry and the conductivity along the $z$-direction is too small to employ as the thermoelectric energy conversion.
The powerfactor PF is defined as PF $=\sigma S^2$.
By comparing the transport properties shown in Figure~\ref{fig:1}(f) with the band structure and DOS shown in Figure~\ref{fig:1}(c)--(d), one can immediately associate the PF peaks with the characteristic band structure and the DOS peaks, as indicated by the three arrows.
It is usually the case in many materials that the PF peak exists near the band edge, where the Seebeck coefficient increases, which corresponds to the PF peak near $\mu=0.6$ eV in our case.
It is also a typical behavior that the second-lowest band edge near the X point, which lies far away from the conduction band bottom, has a very small effect on the PF because of the small Seebeck coefficient. However, it is interesting that the vHs produces the other PF peak near $\mu=1.4$ eV because of its large DOS even though the energy difference between the conduction band bottom and vHs is rather large.
Although the PF peak value near the vHs is a bit smaller than that at the conduction band edge in LaOBiS$_2$, it can be expected that the former peak will be greatly strengthened by making the energy difference between the vHs and conduction band edge smaller. This is one of the basic strategies to enhance the thermoelectric performance in pnictogen-dichalcogenide layered materials, which we shall investigate in further detail in the following subsections.

Before proceeding to the next subsection, we make a few comments on the correspondence between our theoretical results and experimental observations.
Some experimental studies have reported the thermoelectric properties of LaOBiS$_2$ without the F substitution~\cite{LaOFBiS2_thermo,LaOFBiS2_Hall,LaOBiSSe_3}.
Here, we note that LaOBiS$_2$ even without the F substitution is known to possess a small number of electron carriers.
In Ref.~\onlinecite{LaOFBiS2_thermo}, $S\sim -70\ \mu$V K$^{-1}$ was reported, and the carrier concentration was not measured.
In Refs.~\onlinecite{LaOFBiS2_Hall} and \onlinecite{LaOBiSSe_3}, $S\sim -120\ \mu$V K$^{-1}$ with the electron carrier number $n\sim 2\times 10^{-3}$ electron f.u.$^{-1}$ ($1.6\times 10^{19}$ electron cm$^{-3}$) and $S\sim -60\ \mu$V K$^{-1}$ with $n\sim 4\times 10^{-2}$ electron f.u.$^{-1}$ ($3\times 10^{20}$ electron cm$^{-3}$) were reported based on the measurement of the Hall coefficient, respectively.
However, our calculation gives $(S_{xx}, S_{zz})\sim (-280,-220)$ and $(-40,-30)\ \mu$V K$^{-1}$ for $n\sim 2\times 10^{-3}$ and $4\times 10^{-2}$ electron f.u.$^{-1}$, respectively.
Although these calculation results were obtained for the tetragonal structure, we also obtained similar values, $(S_{xx}, S_{yy}, S_{zz})\sim (-250,-240,-200)$ and $(-60,-50,-50)\ \mu$V K$^{-1}$ for $n\sim 2\times 10^{-3}$ and $4\times 10^{-2}$ electron f.u.$^{-1}$, respectively, for the monoclinic structure (see the APPENDIX for details of the calculation for the monoclinic structure).
The discrepancy between the experiments and our calculations can be understood within the large error bar in the Hall coefficient. In addition, all the samples are polycrystals; therefore, a rough correspondence exists between theory and experiment.

A more interesting issue is whether the experiments observe the theoretical PF peak near the doping rate of 0.4 electron f.u.$^{-1}$, which corresponds to the vHs of the DOS.
The thermoelectric properties of LaO$_{1-x}$F$_x$BiS$_2$ were measured in one of the above experimental studies~\cite{LaOFBiS2_thermo} with $x=0$, 0.05, 0.25, and 0.5. However, because the samples with $x=0.25$ and $x=0.5$ exhibit very similar values both for the Seebeck coefficient and electrical conductivity, it appears that the actual doping rate does not reach 40\% of electron doping, which should give the aforementioned theoretical PF peak. In fact, the observed PF monotonically decreases upon increasing the value of $x$ in that reference.
However, because there is a very good agreement between the theoretical band structure and that obtained in the ARPES expreiments~\cite{LaOFBiS2_gap1,NdARPES}, the interesting reascending of PF, as observed in Figure~\ref{fig:1}(f) (or the similar behavior for the monoclinic structure shown in the APPENDIX) is expected to be observed provided that a further amount of electron carriers would be successfully doped.

\subsection{Key factors for the thermoelectric performance\label{sec:key}}

Based on the observation made in the previous subsection for LaOBiS$_2$, one can notice some {\it key factors} that determine its thermoelectric performance.
In this subsection, we carefully investigate each of these factors to pursue possible enhancement of its thermoelectric performance.

\subsubsection{Spin--orbit coupling\label{sec:soc}}

\begin{figure}
\begin{center}
\includegraphics[width=8.5 cm]{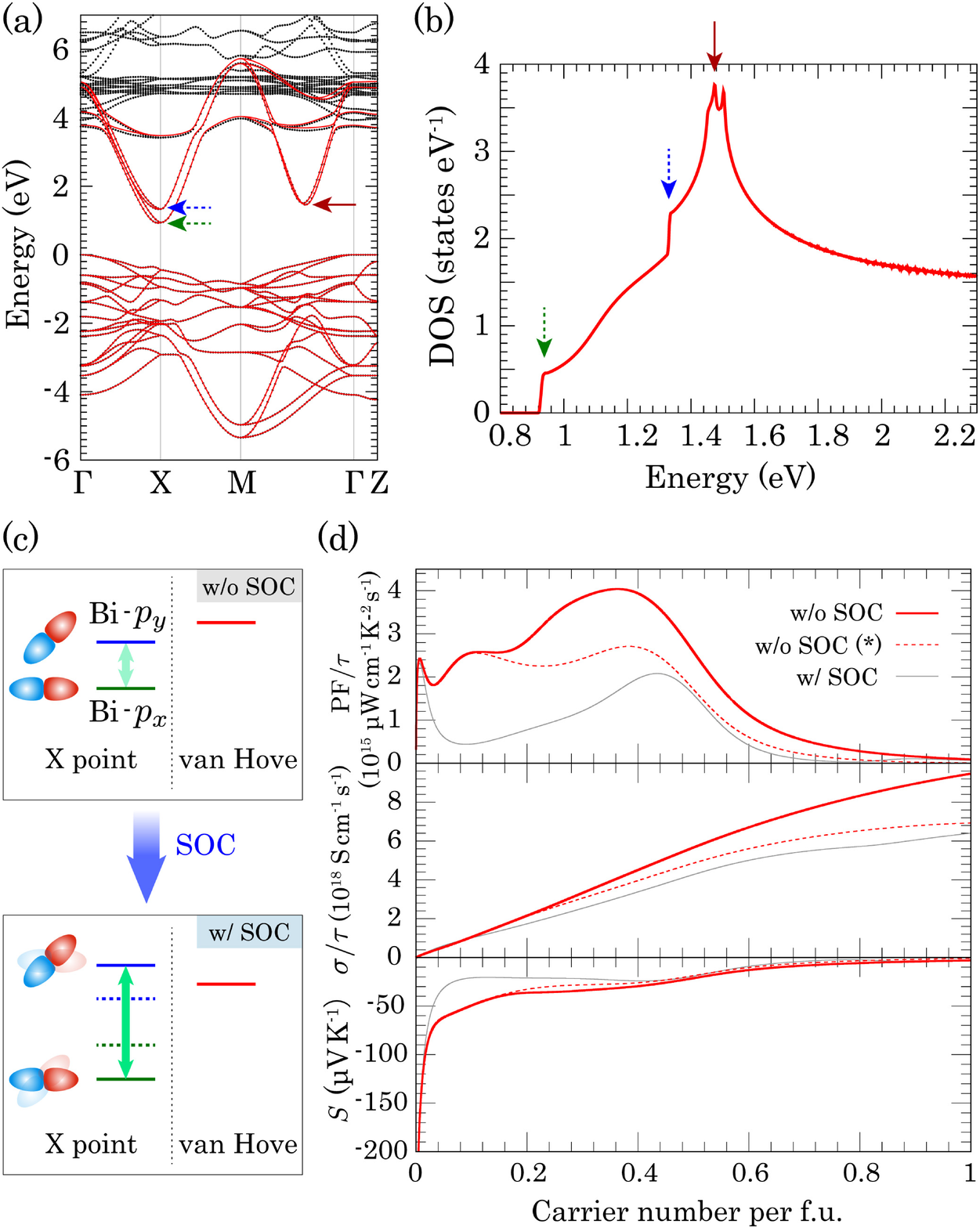}
\caption{(a) Electronic band structures obtained using first-principles calculations (black dotted lines) and model calculations (red solid lines), and (b) DOS for LaOBiS$_2$ without the SOC. The three arrows in (b) correspond to those in (a). (c) Schematic figure showing the effect of the SOC on the energy levels indicated by the three arrows in (a). (d) Transport properties calculated with and without the SOC. The definition of `w/o SOC (*)' calculation is presented in the main text.}
\label{fig:2}
\end{center}
\end{figure}

One of the key factors is the SOC.
{Figure~\ref{fig:2}(a)--(b) presents the band structure and DOS for LaOBiS$_2$ without the SOC.
By comparing them with those obtained with the SOC as shown in Figure~\ref{fig:1}(c)--(d), two important changes are observed.
One is the decrease of the energy difference between the conduction band bottom near the X point and the vHs, from 0.8 to 0.5 eV.
The other one is the considerable reduction of the band splitting between the lowest and second-lowest conduction bands at the X point, from 1.2 to 0.4 eV.
Both of these features are advantageous for the thermoelectric performance because they augment the DOS in the low-energy region of the conduction bands.

The physics behind these characteristic changes can be naturally understood using the schematic figure in Figure~\ref{fig:2}(c).
First, it is known that the Bi-$p_x$ and Bi-$p_y$ orbitals mainly constitute the Bloch states in the lowest and second-lowest conduction bands at the X point, respectively, when the SOC is switched off~\cite{BiS2_HP}.
However, the SOC introduces the coupling between these two states and then enlarges the energy splitting between them.
However, at the $k$-points that correspond to the vHs, the lowest conduction band lies far from the second-lowest one with respect to their energy levels, as observed in Figure~\ref{fig:2}(a).
This large separation reduces the effect of the SOC, and therefore, the energy level of the vHs is expected to change less than those at the X point.
As a result, the relative energy level of the vHs to the lowest energy level at the X point increases with the SOC.

Figure~\ref{fig:2}(d) presents the thermoelectric properties with and without the SOC for LaOBiS$_2$.
To demonstrate the effect of the second-lowest conduction band at the X point on the transport properties, we also calculated the transport quantities using only the lowest conduction bands (to be more precise, the lowest two conduction bands with the small bilayer splitting), which we denote as `w/o SOC (*)' in the figure.
In this calculation, we used the chemical potential $\mu (T)$ determined by the full inclusion of all the band dispersions, which we denote as `w/o SOC' in the figure, to make comparison between them.
As expected from the above discussion, the PF dramatically increases in a wide range of the carrier doping rate when the SOC is switched off.
Because the `w/o SOC', `w/o SOC (*)', and `w/SOC' results exhibit sizable differences, the energy lowering of both the vHs and second-lowest band edge near the X point are found to contribute to the thermoelectric performance. In particular, the second-lowest band begins to enhance the PF above the carrier numbers of approximately 0.2 electron f.u.$^{-1}$, which can be verified by the difference between the `w/o SOC' and `w/o SOC (*)' results.
Although the vHs is higher than the second-lowest band edge near the X point, the energy lowering of the vHs also contributes to the PF for much lower carrier concentration, as demonstrated by the difference between the `w/o SOC' and `w/SOC' results. This finding is observed because the energy lowering of the vHs increases the DOS between the vHs and the lowest band edge, as evidenced in Figures~\ref{fig:1}(d) and \ref{fig:2}(b).

\subsubsection{Interatomic hopping amplitudes\label{sec:hop}}

\begin{figure}
\begin{center}
\includegraphics[width=8.5 cm]{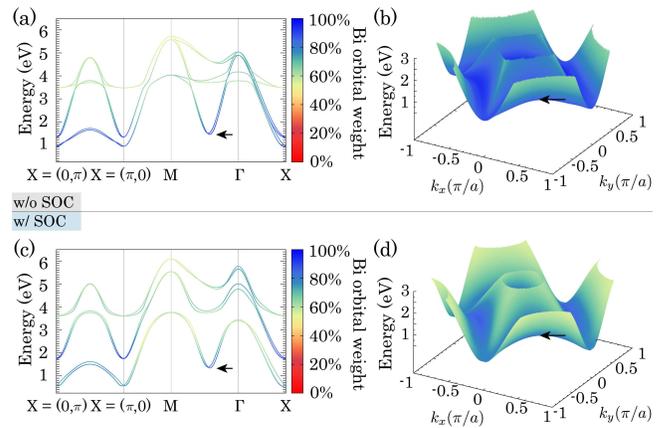}
\caption{Band structure colored by Bi orbital weight (a) along some $\bm{k}$-paths and (b) on the $k_z=0$ plane, which were calculated without the SOC. In (b), only the lowest conduction band is depicted. (c),(d) The same plots for the calculation with the SOC.}
\label{fig:3}
\end{center}
\end{figure}

Figure~\ref{fig:3}(a) and (c) presents the Bi orbital weight on the band structure of LaOBiS$_2$ calculated using our tight-binding model without and with the SOC, respectively.
We can see that both the vHs point indicated by the arrows and the two lowest conduction bands at the X point have a very large Bi orbital weight regardless of the inclusion of the SOC.
This situation can be understood based on the following reasons.
(i) Some of the present authors and their coworker have shown that the in-plane hybridization among Bi and S orbitals is forbidden at the X point by the crystal symmetry~\cite{BiS2_HP}. Although the inter-BiS$_2$-layer coupling allows the hybridization between the Bi and S states, the small bilayer coupling $\sim \mathcal{O}$(0.1 eV) compared with the onsite energy difference between the Bi and S atomic orbitals $\sim \mathcal{O}$(1 eV) results in a very small weight of the S orbitals there.
(ii) The very large band dispersion along the M--$\Gamma$ line, as observed in Figure~\ref{fig:3}(a) and (c), originates from the strong hybridization between the in-plane nearest-neighboring Bi and S atomic orbitals ($t^{\mathrm{Bi-S}}_{\sigma,\pi}$ shown in Figure~\ref{fig:1}(b)). Here, stronger hybridization for the Bloch states at each $\bm{k}$-point results in a higher energy level of the conduction band, which is a general consequence of the bonding--antibonding splitting.
Therefore, the vHs, which corresponds to the lowest energy along the M--$\Gamma$ line, should have (almost) the smallest S weight along this line.

Because of this large Bi weight for the two lowest conduction bands at the X point and vHs, the Bi--Bi hopping amplitudes play a dominant role in determining their energy levels. In particular, when the Bi--Bi hopping amplitudes become smaller, the energy differences among these states should also decrease because they all approach the Bi onsite energy. Therefore, to improve the thermoelectric performance by increasing the DOS around the band edge, smaller Bi--Bi hopping amplitudes are advantageous.

As shown in Figure~\ref{fig:3}(b) and (d), it is quite impressive that the band dispersion exhibits a strong quasi-one-dimensionality because of the strong Bi--S hopping amplitudes, which induce the sharp band dispersion along the lines $k_x=\pm k_y$ (i.e., along the M--$\Gamma$ line), and the relatively weak Bi--Bi hopping amplitudes, which make the band dispersion along the vertical direction, i.e., the lines $|k_x| + |k_y| = \pi$, less dispersive~\cite{Usui}.
The quasi-one-dimensional band structure reconciles the large DOS and high group velocity and, therefore, is very advantageous for the thermoelectric performance~\cite{Hicksone,Hicksone2,usuione}. This concept is similar to that of the pudding-mold-shaped band structure~\cite{pudding}.
To enhance the one-dimensionality of the band dispersion, larger in-plane Bi--S hopping amplitudes are advantageous.

\subsubsection{Onsite energy difference}

\begin{figure}
\begin{center}
\includegraphics[width=8.5 cm]{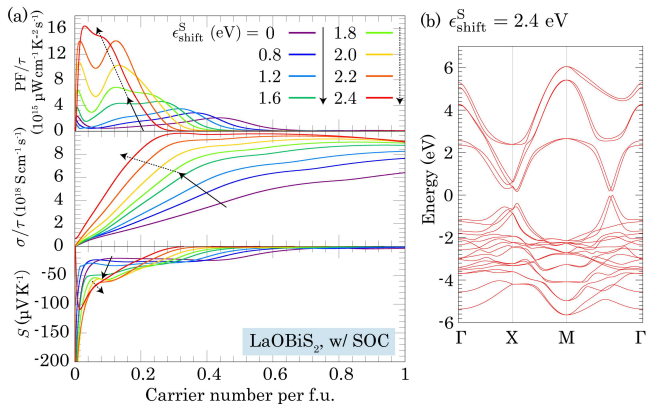}
\caption{(a) Variation of the transport quantities against the change of the in-plane S-$p_{x,y}$ onsite energy shift, $\epsilon^{\mathrm{S}}_{\mathrm{shift}}$, for LaOBiS$_2$ with the SOC. (b) Band structure calculated with the tight-binding model with $\epsilon^{\mathrm{S}}_{\mathrm{shift}} = 2.4$ eV.}
\label{fig:4}
\end{center}
\end{figure}

The onsite energy difference between the Bi-$p_{x,y}$ and in-plane S-$p_{x,y}$ orbitals should also have a large effect on the thermoelectric property.
To demonstrate this effect, we calculated the transport properties by shifting the in-plane S-$p_{x,y}$ onsite energies by $\epsilon^{\mathrm{S}}_{\mathrm{shift}}$ for LaOBiS$_2$ with the SOC, as shown in Figure~\ref{fig:4}(a).
Here, without the onsite energy shift, the onsite energy difference between the Bi-$p_{x,y}$ and in-plane S-$p_{x,y}$ orbitals is 2.1 eV.
We can see that the PF peak value considerably increases by increasing $\epsilon^{\mathrm{S}}_{\mathrm{shift}}$.
In Figure~\ref{fig:4}(b), we show the band structure calculated with the tight-binding model with $\epsilon^{\mathrm{S}}_{\mathrm{shift}} = 2.4$ eV, where the onsite energy difference becomes much smaller than the original value.
As clearly observed, the band gap becomes very small $\sim 0.2$ eV, and the conduction band dispersions are sharpened compared with the original one shown in Figure~\ref{fig:1}(c).
Because hybridization between the valence and conduction bands opens a relatively large gap at the X point, the onsite energy shift also brings the vHs close to the conduction band bottom.
These effects enhance PF.
Because we applied a somewhat artificial change of the model parameters here, the details of the band structure change would be different in real compounds, where the onsite energy difference is controlled by elemental substitution. Nevertheless, from the monotonic increase of PF observed in Figure~\ref{fig:4}(a), we can expect that a decrease of the onsite energy difference will in general largely improve the thermoelectric performance.

\subsubsection{Promising recipe for materials design}

Here, we summarize the key factors that can improve the thermoelectric performance of LaOBiS$_2$:
(1) small SOC, (2) small Bi--Bi and large Bi--S hopping amplitudes, and (3) small onsite energy difference between the Bi-$p_{x,y}$ and in-plane S-$p_{x,y}$ orbitals.
Surprisingly and fortunately, these observations lead us to {\it a rather simple strategy for materials design}: replacing Bi and S atoms with lighter and heavier elements, respectively, and hence placing them closer in the periodic table.
In fact, when one replaces Bi atoms with lighter elements, the SOC is trivially weakened; in addition, the Bi--Bi hopping amplitudes will be reduced because the spread of the Bi atomic orbitals becomes small. By replacing S atoms with heavier elements, we can further reduce the Bi--Bi hopping by increasing their distance as a result of the enlarged spread of the S atomic orbitals.
How these changes affect the Bi--S hopping amplitudes is not trivial, and therefore, we shall verify this point later in this paper.
Moreover, the Bi (S) onsite energy decreases (increases) by the aforementioned atomic replacement within the same row in the periodic table, and then, the onsite energy difference will be reduced.

\subsection{Possible large PF and $ZT$ by elemental substitution\label{sec:material}}

\begin{figure*}
\begin{center}
\includegraphics[width=17 cm]{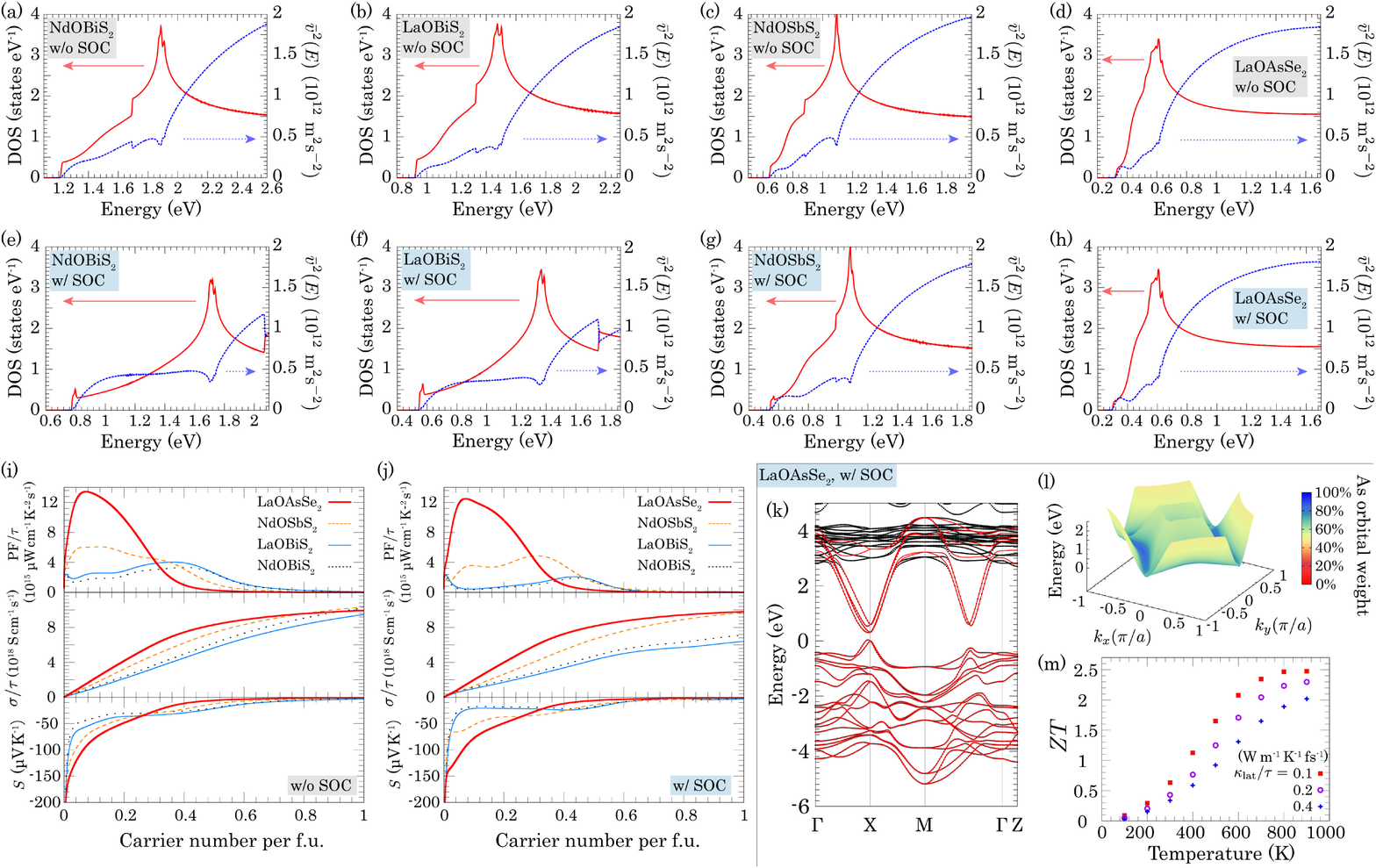}
\caption{DOS and $\bar{v}^2 (E)$ (Equation~(\ref{eq:v2})) for (a) NdOBiS$_2$, (b) LaOBiS$_2$, (c) NdOSbS$_2$, and (d) LaOAsSe$_2$ calculated without the SOC. (e)--(h) The same quantities calculated with the SOC. The transport properties of these four compounds calculated (i) without and (j) with the SOC. (k) First-principles band structure, (l) ($k_x$, $k_y$, $E$) plot for the lowest conduction band on the $k_z=0$ plane, and (m) estimated $ZT$ values for LaOAsSe$_2$ calculated with the SOC. The As orbital weight is indicated by the color scale in (l).}
\label{fig:5}
\end{center}
\end{figure*}

\begin{table}
\begin{center}
\begin{tabular}{c c c c c c c c c}
\hline \hline
(eV) &  SOC & $\Delta$ & $t^{\mathrm{Bi}-\mathrm{S}}_{\sigma}$ &$t^{\mathrm{Bi}-\mathrm{S}}_{\pi}$ & $t^{\mathrm{Bi}-\mathrm{Bi}}_{\sigma}$ & $t^{\mathrm{Bi}-\mathrm{Bi}}_{\pi}$ & $E^{\mathrm{split}}_{\mathrm{X}}$ & $E^{\mathrm{diff}}_{\mathrm{vH}}$\\
\hline
NdOBiS$_2$ & no & 2.24 & 2.23 & -0.50 & 0.28 & 0.13 & 0.5 & 0.6 \\
& yes & 2.20 & 2.23 & -0.50 & 0.27 & 0.13 & 1.3 & 0.9 \\
LaOBiS$_2$ & no & 2.16 & 2.15 & -0.47 & 0.26 & 0.12 & 0.4 & 0.5 \\
& yes & 2.12 & 2.14 & -0.47 & 0.27 & 0.12 & 1.2 & 0.8 \\
NdOSbS$_2$ & no & 1.83 & 2.18 & -0.48 & 0.21 & 0.13 & 0.2 & 0.4 \\
& yes & 1.82 & 2.18 & -0.48 & 0.21 & 0.13 & 0.4 & 0.5 \\
LaOAsSe$_2$ & no & 0.78 & 1.92 & -0.40 & 0.11 & 0.07 & 0.2 & 0.2 \\
& yes & 0.78 & 1.92 & -0.40 & 0.11 & 0.07 & 0.3 & 0.2 \\
\hline \hline
\end{tabular}
\end{center}
\caption{\label{table:hop}
Hopping parameters defined in Figure 1(b) and the onsite energy difference between the Bi-$p_{x,y}$ and in-plane S-$p_{x,y}$ orbitals, $\Delta$. Whereas the Bi and S atoms are replaced with other atoms for some compounds, we use the notation `Bi' and `S' here for simplicity. The row `SOC' indicates whether the SOC is included. The energy difference between the lowest and second-lowest conduction bands at the X point, $E^{\mathrm{split}}_{\mathrm{X}}$, and that between the lowest conduction bands at the X point and at the vHs, $E^{\mathrm{diff}}_{\mathrm{vH}}$, are also presented.}
\end{table}

To investigate the effects of elemental substitution, we calculated and compared the thermoelectric properties of NdOBiS$_2$, LaOBiS$_2$, NdOSbS$_2$, and LaOAsSe$_2$.
Figure~\ref{fig:5}(a)--(h) presents the DOS and $\bar{v}^2 (E)$ for these four compounds with and without the SOC, where $\bar{v}^2 (E)$ is the averaged value of the square of the group velocity, as defined by
\begin{equation}
\bar{v}^2 (E) = \frac{\sum_{n,\bm{k}} v^2_{n,\bm{k}} \delta (E- \epsilon_{n,\bm{k}})}{\sum_{n,\bm{k}} \delta (E - \epsilon_{n,\bm{k}})},\label{eq:v2}
\end{equation}
with $n$ and $\bm{k}$ being the band and $\bm{k}$-point indices, respectively.
Comparison of Figure~\ref{fig:5}(a)--(h) reveals that the DOS peak is closer to the band edge in the order of LaOAsSe$_2$, NdOSbS$_2$, LaOBiS$_2$, and NdOBiS$_2$. Such DOS enhancement near the band edge coincides with the decrease of the Bi--Bi (Sb--Sb or As--As) hopping amplitudes, as observed in Table~\ref{table:hop}, which is precisely the expected behavior discussed in Section~\ref{sec:hop}. The difference between NdOBiS$_2$ and LaOBiS$_2$ originates from the lattice constant, where the smaller lattice constants in NdOBiS$_2$ result in a slight increase of the Bi--Bi hopping amplitudes.
Whereas NdOBiS$_2$, LaOBiS$_2$, and NdOSbS$_2$ have similar Bi--S (Sb--S) hopping amplitudes, LaOAsSe$_2$ has slightly smaller ones, as observed in Table~\ref{table:hop}. However, the decrease of the hopping amplitudes between Bi (As) atoms, $t^{\mathrm{Bi}-\mathrm{Bi}}$, is much larger with respect to the change ratio. Because of this feature, $\bar{v}^2 (E)$ near the band edge has roughly similar values for these four compounds when the SOC is switched off, as observed in Figure~\ref{fig:5}(a)--(d)~\cite{SOC_note}. Because the group velocity usually decreases when the DOS increases, the coexistence of the high group velocity and large DOS in LaOAsSe$_2$ is very advantageous for the thermoelectric performance, as discussed in Section~\ref{sec:hop}. This advantage originates from the quasi-one-dimensional band structure, as shown in Figure~\ref{fig:5}(k)--(l).
The effect of the SOC becomes weaker with the atomic displacement from Bi to Sb and As, although it still retains a sizable strength for Sb, as observed in Figure~\ref{fig:5}(c) and (g).

Figure~\ref{fig:5}(i)--(j) shows the calculated thermoelectric performance for these four compounds with and without the SOC.
The prominent feature here is the distinguished PF/$\tau$ of LaOAsSe$_2$, $12\times 10^{15}\ \mu$W cm$^{-1}$ K$^{-2}$ s$^{-1}$ at its peak value, which is six times as large as that of LaOBiS$_2$, as shown in Figure~\ref{fig:5}(j).
This considerable enhancement amounts to PF $\sim 60$--$120\ \mu$W cm$^{-1}$ K$^{-2}$ at room temperature when one assumes a typical relaxation time $\tau \sim$ 5--10 fs.
In addition, we also estimated the $ZT$ values of LaOAsSe$_2$ at several temperatures, as presented in Figure~\ref{fig:5}(m).
For evaluation of $ZT$ under the constant relaxation-time approximation, there is only one unknown parameter, $\kappa_{\mathrm{lat}} / \tau$, because $\tau$ in PF and the electrical thermal conductivity $\kappa_{\mathrm{el}}$ are canceled. Some experimental studies have provided estimations of the lattice thermal conductivity $\kappa_{\mathrm{lat}}$, $\sim$ 1 W m$^{-1}$ K$^{-1}$ at 300--700 K for LaOBiSSe~\cite{LaOBiSSe_HP} and $\sim$ 2 W m$^{-1}$ K$^{-1}$ at 50--300 K for LaOBiS$_2$~\cite{LaOFBiS2_Hall}.
Based on these values, we calculated the $ZT$ value using three ratios for $\kappa_{\mathrm{lat}} / \tau$: 0.1, 0.2, and 0.4 W m$^{-1}$ K$^{-1}$ fs$^{-1}$.
We note that the experimental thermal conductivities were obtained for (weakly oriented) polycrystalline samples and thus can be underestimated as the in-plane thermal conductivity for single crystals or highly oriented polycrystals. This is another reason why we tried several $\kappa_{\mathrm{lat}} / \tau$ values in estimating the $ZT$ value here.
For all of these three ratios of $\kappa_{\mathrm{lat}} / \tau$, LaOAsSe$_2$ exhibits very high $ZT$ values, which become $\sim$ 0.5 near room temperature and reach approximately two in the high-temperature region. 
Note that $ZT$ is maximized at lower doping concentration than the PF peak because of the increase of the electronic thermal conductivity by carrier doping.
For example, when one assumes $\kappa_{\mathrm{lat}}=1$ W m$^{-1}$ K$^{-1}$, $\tau = 5$ fs (i.e., $\kappa_{\mathrm{lat}} / \tau = 0.2$ W m$^{-1}$ K$^{-1}$ fs$^{-1}$), and $T=$ 300 K,
the electronic thermal conductivity $\kappa_{\mathrm{el}}$ reaches $\sim 5$ W m$^{-1}$ K$^{-1}$ for the PF peak at 0.072 electron f.u.$^{-1}$ ($6\times 10^{20}$ electron cm$^{-3}$), which is much larger than the lattice thermal conductivity, whereas $ZT$ is maximized at 0.034 electron f.u.$^{-1}$ ($3\times 10^{20}$ electron cm$^{-3}$).

Although the enhancement of PF is not as large as that in LaOAsSe$_2$, NdOSbS$_2$ also appears to be a promising material as the PF is much larger than that of LaOBiS$_2$ for a wide range of carrier concentrations. We should note that NdOSbS$_2$ has already been synthesized~\cite{NdOSbS2_strct}, although the carrier doping and measurement of its thermoelectric performance have not yet been reported in experiments. Further experimental study for Sb and As compounds is strongly awaited.
We note that for LaOAsSe$_2$, replacement only of in-plane S atoms should suffice for synthesis because the thermoelectric property is governed by the in-plane atoms.

From the small difference of the thermoelectric performance between NdOBiS$_2$ and LaOBiS$_2$, we can conclude that the effect of the blocking layer is not so large, at least when compared with the effect of the pnictogen or chalcogen substitution. The selection of the blocking layer might be, however, rather crucial for synthesizing intended compounds or stabilizing the tetragonal structure we assumed in this study (see the APPENDIX for the effects of the symmetry reduction), or perform carrier doping by F substitution.
In this sense, the large degrees of freedom for the blocking layer might be very helpful for the experimental exploration of high-performance materials.

\subsection{Relevance to Dirac dispersion\label{sec:disc}}

The onsite energy difference of LaOAsSe$_2$ is three times as small as that for LaOBiS$_2$, as shown in Table~\ref{table:hop}.
This reduction was expected; however, by examining the band structure and DOS in Figure~\ref{fig:5}(d), (h), and (k), which reveal a band gap of approximately 0.3 eV regardless of the presence of the SOC, a further reduction of the onsite energy difference is expected to even further enhance the thermoelectric performance.
Interestingly, the `limit' along this line leads to Dirac cones with very strong anisotropy, similar to the observations for $A$MnBi$_2$ ($A$ = Sr, Ca, etc.)~\cite{AMnBi2_1,AMnBi2_2,AMnBi2_3,AMnBi2_4,AMnBi2_5,AMnBi2_6,AMnBi2_7,AMnBi2_8,AMnBi2_9,AMnBi2_10,AMnBi2_11} compounds, where the Bi atoms constitute the square-lattice network.
Indeed, in $A$MnBi2, it was shown in calculations that the gap of the Dirac cone closes when the SOC is switched off~\cite{AMnBi2_1,AMnBi2_2} because the onsite energy difference, which exists in LaOBiS$_2$, vanishes here~\cite{note_Dirac}.
It is also noteworthy that Ta$_4$SiTe$_4$ with one-dimensional Dirac cones was recently observed to exhibit a very large PF of $\sim$ 170 $\mu$W cm$^{-1}$ K$^{-1}$ at 220--280 K~\cite{Ta4SiTe4}.
In fact, we can get a glimpse of the gapped Dirac-like band dispersions near the X point and along the $\Gamma$--M line for LaOAsSe$_2$ and for LaOBiS$_2$ with the onsite energy shift in Figures~\ref{fig:5}(k) and \ref{fig:4}(b), respectively.
Because transport properties including the thermoelectric performance of the Dirac and Weyl fermions have attracted much attention~\cite{Diracs}, the relation between pnictogen-dichalcogenide layered compounds and anisotropic or one-dimensional Dirac cones is intriguing.
In fact, based on the observation that a sharp dispersion (i.e.,~a large group velocity) is favorable for the thermoelectric performance in the one-dimensional band structure~\cite{usuione}, the Dirac or gapped-Dirac band dispersion is desirable because of the large group velocity near the band edge.
We also note that the gap opening for the Dirac cone is important to prevent deterioration of the thermoelectric performance by cancelation in the Seebeck coefficient between the contribution from the electron and hole carrier transport~\cite{gapDirac1,gapDirac2,gapDirac3}.

Along this line, we note that our target materials exhibit favorable thermoelectric performance both for the $x$ and $y$ directions, which is not the case for real one-dimensional materials such as Ta$_4$SiTe$_4$. This characteristic is of great importance for technological applications because not only single crystals but also polycrystals with high orientation only along the $z$-direction are expected to work well because of this two-dimensionality, which enables two-dimensional conduction. In contrast, real one-dimensional materials that exhibit good thermoelectric performance only for one direction would not work well in polycrystals.
Therefore, we arrive at the novel concept of ``{\it quasi-one-dimensional gapped-Dirac cones in two-dimensional materials}'', which is considered one of the ideal conditions for thermoelectric materials. 
 
\section{Conclusion\label{sec:sum}}
In this study, we investigated the thermoelectric performance of LaOBiS$_2$ and uncovered a theoretical strategy for improving its performance.
This strategy works quite well, as demonstrated by the comparison of four compounds: NdOBiS$_2$, LaOBiS$_2$, NdOSbS$_2$, and LaOAsSe$_2$.
The atomic replacements of Bi with Sb and As and that of S with Se largely improve the thermoelectric performance, and in particular, LaOAsSe$_2$ exhibits a remarkably high PF and $ZT$. This high performance is attributed to the quasi-one-dimensional gapped Dirac-like band dispersion, where the large DOS and high group velocity can coexist. Here, the square-lattice network offers an ideal arena for such a favorable band structure.

\acknowledgments
We appreciate fruitful discussion with Yoshikazu Mizuguchi, Yosuke Goto, Chul-Ho Lee, and Hideaki Sakai.
This study was supported by JSPS KAKENHI Grant (Nos.~JP17H05481, JP26610101, and JP17K14108) and JST CREST (No.~JPMJCR16Q6), Japan.

\section*{APPENDIX: Electronic properties of the monoclinic structure of LaOBiS$_2$}
Figure~\ref{fig:app} presents the calculated band structure, DOS, and transport properties of LaOBiS$_2$ with the monoclinic structure (space group: $P2_1$/$m$). 
All the calculations presented in this figure include the SOC. We used the experimental crystal structure ($a=4.0769(4)$ \AA, $b=4.0618(3)$ \AA, $c=13.885(2)$ \AA, $\beta=90.12(2)^\circ$) taken from Ref.~\onlinecite{SymLow}. For our transport calculations, the $x$ and $y$ axes were taken to be parallel to the $a$ and $b$ axes, respectively. Here, we show the diagonal components of the transport quantities for the in-plane directions.

In Figure~\ref{fig:app}(a) and (b), we can see that the DOS peaks are split by the symmetry reduction in the monoclinic structure.
This split is caused by the inequivalency between the $x$ and $y$ directions, as mentioned above.
As a result, the PF peak values presented in Figure~\ref{fig:app}(c) decrease. Based on this observation, the symmetry reduction is not desirable for high-thermoelectric performance, which is another important guiding principle for the materials search among pnictogen-dichalcogenide layered compounds.

\begin{figure*}
\begin{center}
\includegraphics[width=17 cm]{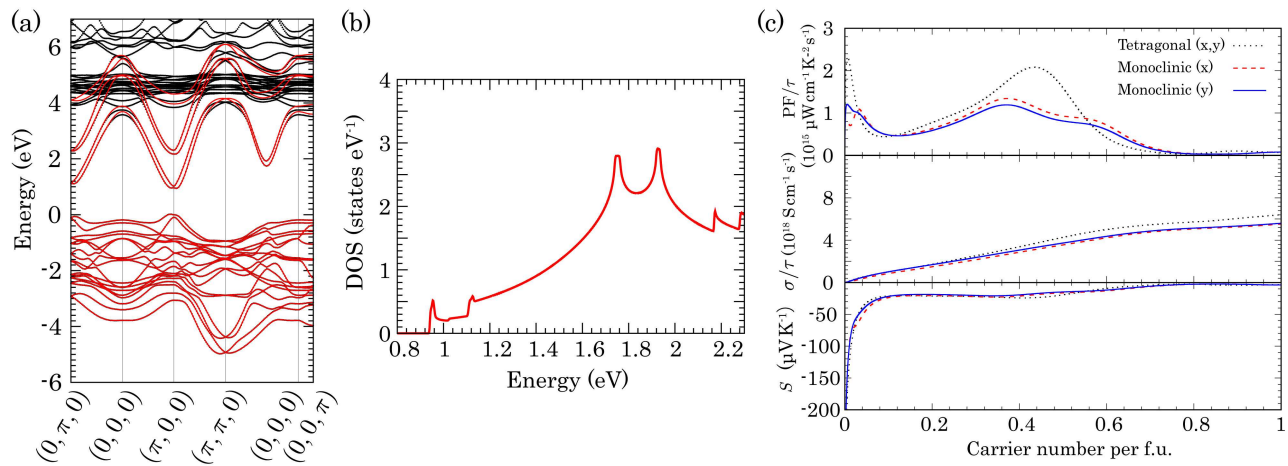}
\caption{(a) Electronic band structure obtained from first-principles calculations (black dotted lines) and model calculations (red solid lines) for the monoclinic structure of LaOBiS$_2$. The $\bm{k}$-points are presented with the fractional coordinate multiplied by $2\pi$. (b) DOS for the monoclinic LaOBiS$_2$ and (c) transport properties calculated using the tight-binding model. In (c), the diagonal components of each tensor quantities are presented. We note that the $x$- and $y$-directions are not equivalent for the monoclinic structure, unlike for the tetragonal one. All the calculations in this figure were performed including the SOC.}
\label{fig:app}
\end{center}
\end{figure*}


\begin{thebibliography}{999}
\bibitem{note_NdOBiS2_strct} Whereas the crystal structure shown in Ref.~\cite{NdOBiS2_strct} corresponds to F-substituted NdOBiS$_2$, Nd$_{0.98\pm 0.06}$O$_{0.7\pm 0.1}$F$_{0.3\pm 0.1}$Bi$_{0.98\pm 0.04}$, the effect of the rigid band approximation against the carrier doping through the change of the lattice constants is shown to be small in Sec. 3.3 in the main text.
\bibitem{note:gap} For the monoclinic structure, the band gap calculated with the mBJ potential is 0.9 eV, as shown in the APPENDIX. The difference mainly originates from the valence bands near the $\Gamma$ point, which become the valence band maximum for the tetragonal structure but not for the monoclinic one. The direct band gap near the X $=(\pi,0,0)$ point is similar near 0.9 eV for both structures.
\bibitem{SOC_note} Although the SOC widens the band width, as shown in Figure~\ref{fig:2}(c), and hence increases the group velocity in LaOBiS$_2$ and NdOBiS$_2$, as shown in Figure~\ref{fig:5}(e)--(f), we have already seen that the SOC deteriorates the thermoelectric performance in Sec.~\ref{sec:soc}.
\bibitem{note_Dirac} In Ref.~\onlinecite{Dirac_SbSe2}, it was theoretically pointed out that LaOSbSe$_2$ thin films can possess multiple Dirac cones under an electric field, while the calculations in that paper were based on GGA, which is known to underestimate the band gap.
\end{thebibliography}
\end{document}